\documentclass[10pt, conference]{IEEEtran}

\usepackage{subcaption}
\usepackage{booktabs}
\usepackage{array}
\usepackage{amsmath,amssymb,amsfonts}
\usepackage{algorithm}
\usepackage{graphicx}
\usepackage{textcomp}
\usepackage{float}
\usepackage{listings}
\usepackage{amsmath,boxedminipage}
\usepackage{hyperref}
\usepackage{url}

\usepackage{xcolor}
\usepackage{balance}

\usepackage[utf8]{inputenc}
\usepackage[noend]{algpseudocode}
\usepackage[utf8]{inputenc}
\usepackage[english]{babel}
\usepackage{xspace}
\usepackage{tabularx}
\usepackage{multirow}
\usepackage{paralist}
\usepackage[skins]{tcolorbox}
\usepackage{enumitem,kantlipsum}

\begin{document}

%\title{Teaching Machines to Read API Documentation for Answering Developer Questions}
%\title{Learning to Recommend Relevant API Documentation for Answering Developer Questions}

\title{An Empirical Study on the Characteristics of Question-Answering Process on Developer Forums}

\author{\IEEEauthorblockN{Yi Li\textsuperscript{1}, Shaohua Wang\textsuperscript{1}, Tien N. Nguyen\textsuperscript{2}, Son Van Nguyen\textsuperscript{2}, Xinyue Ye\textsuperscript{1}, Yan Wang\textsuperscript{3}}
	
	\IEEEauthorblockA{
		\textsuperscript{1}Department of Informatics, New Jersey Institute of Technology, Newark, United States\\
		\{yl622, shaohua.wang, xinyue.ye\}@njit.edu\\
		\textsuperscript{2}Computer Science Department, The University of Texas at Dallas, United States, \{tien.n.nguyen,sonnguyen\}@utdallas.edu\\
		\textsuperscript{3}Central University of Finance and Economics, China, dayanking@gmail.com}\\
	
}
\maketitle

	\vspace{-30pt}
\begin{abstract}

Developer forums are one of the most popular and useful Q\&A
websites on API usages. The analysis of API forums can be a critical
resource for the automated question and answer approaches. In this
paper, we empirically study three API forums including Twitter, eBay,
and AdWords, to investigate the characteristics of question-answering
process. We observe that +60\% of the posts on all three forums were
answered by providing API method names or documentation. +85\% of the
questions were answered by API development teams and the answers from
API development teams drew fewer follow-up questions. Our results
provide empirical evidences for us in a future work to build automated
solutions to answer developer questions on API forums.

\end{abstract}

\maketitle

\vspace{-10pt}
\section{Introduction}
\vspace{-5pt}

%Application Programming Interfaces (APIs) have become the backbone in
%modern software development. Accurately and effectively using APIs
%becomes extremely critical in software
%development~\cite{uddin2017automatic}.  
Recently, developer question
and answering (Q\&A) websites have become popular, critical and
essential on-line resources that developers use to seek for their
solutions on API usages, to share and learn knowledge of using APIs,
and even to make discussions on the design of
APIs~\cite{mamykina2011design}.

\iffalse
\begin{figure}[h]
	\centering
	\includegraphics[width=8cm]{graphs/Graph_1.jpg}
	\caption{A pair of real API function related question and answer from Twitter developer forum. The top is the developer's question and the bottom is the API development team member's answer}
	\label{Fig.1}	
\end{figure}
\fi

Recently, two main types of developer Q\&A websites (DQA) have become
popular. The first type is the general-purpose Q\&A websites, for
example, Stack Overflow~\cite{stackoverflow}), taking any questions
relevant to any APIs. The second type of DQA is the API Q\&A forums
maintained by the libraries' providers, e.g., Twitter~\cite{twitter},
and they accept only the questions relevant to the APIs of specific
libraries.  The main differences between the two DQAs can be
summarized as follows: (1) Typically, an API forum is run by a
library's provider and has the members from the development team to
answer the questions relevant to the APIs of the libraries. 
%The empirical investigation in this paper suggests that about 87\% of the questions that have answers on the API forums are answered by the API development teams. 
Developers tend to ask API-specific questions on
API forums. However, StackOverflow tends to deem the valid questions
yet specific to a particular library as off-topic
questions~\cite{squire2015should}. The API development teams on API
forums can offer fast and right-to-the-point responses to the API
specific questions~\cite{venkatesh2016client}; and (2) Typically, the
general-purpose DQA provides incentives%, %e.g., badges and a voting system, 
to improve the credibility of the responders and their public
answers~\cite{wang2018users}. A question or an answer can be modified
multiple times on SO, while the API Q\&A forums often do not allow
developers to modify others' questions or answers. Due to those
major differences, it is necessary to help API development teams
answer more questions, and provide high-quality and right-to-the-point
answers on API~forums.

Extensive
research~\cite{zhang2019empirical,calefato2018ask,calefato2019empirical,li2018learning}
has been devoted to studying StackOverflow (SO), one of the most
popular DQA websites. However, despite the importance of API forums,
little research has been focused on library-specific forums. In this
paper, we set out to investigate the process of question-answering on
such forums. We empirically studied three popular API Q\&A forums,
Twitter~\cite{twitter}, eBay~\cite{ebay}, and Google
AdWords~\cite{AdWords}, to answer the following research
questions:

\noindent\textbf{RQ1. How are the questions answered?}

In this RQ, we want to study how a question is answered by developers
on an API forum.  Our results indicate that majority of the questions
were answered with provided API method names (or sometimes links to
API documentation).

\noindent\textbf{RQ2. Who answer the questions?}

Similar to the general purpose DQA websites, any developer can answer
a question on API forums.  However, we observe that the majority of
the studied questions were answered by API development teams.

\noindent\textbf{RQ3. What is the quality of answers?}

Our further analysis of the answered questions on API forums shows
that the answers from API development teams have drawn fewer follow-up
questions than the ones answered by other developers.

\section{Empirical Study Design}
\vspace{-5pt}
Our overall goal is to understand the process of question-answering on
the library-specific forums.

%\vspace{2pt}
\noindent\textbf{Data Collection and Processing.} We conducted an
empirical study on three popular web-based Q\&A forums, including
Twitter, eBay, and AdWords, to investigate the basics of developer API
Q\&A forums and to motivate our study using our findings. We built a
tool to crawl all of the questions and their answers from each of the
aforementioned developer Q\&A forums (last access in April
2018). Table~\ref{Questions_dataset} shows that over 50\% of the
questions on each forum were not answered.

%Firstly, we did a statistic analysis on three API developer forums include twitter, eBay, and AdWords since the April 30th 2019. The detailed information of all questions in these three data sets are shown in figure \ref{Questions_dataset}

\begin{table}[h]
	\caption{Statistics of Each library-specific Q\&A Forum.}
	\vspace{-10pt}
	\begin{center}
		%\vspace{-10pt}
		\renewcommand{\arraystretch}{1}
		\begin{tabular}{p{3.8cm}|p{1cm}<{\centering}p{1cm}<{\centering}p{1cm}<{\centering}}
			%\hline
			\hline
			 & Twitter  & eBay & AdWords \\
			\hline
			Total \# of questions	&16,874	&6,204	&23,731\\
			\# of questions with answers	&8,910	&3,524	&12,364\\
			\# of questions without answers	&7,964	&2,680	&11,367\\
			\% of questions without answers &52.8\%	&56.8\%	&52.1\%\\
			\hline
		\end{tabular}
		\label{Questions_dataset}
		\vspace{-20pt}
	\end{center}
\end{table}

%\vspace{2pt}
\noindent\textbf{Analysis Approach for RQs.} We further analyze the
answered questions of each forum to study who answer questions and how
questions were answered. Using the confidence level 95\% with an
interval 5\%, we randomly selected 368, 358, 374 questions from 8,910,
5,231, and 14,245 questions having answers on Twitter, eBay, and
AdWords, respectively. For each selected question, we manually studied
the question and its answers to classify the question based on how it
was answered.

%Through our study, we make the following observations:

There can be many metrics for measuring answer quality. For
simplicity, in this short paper, we use the number of follow-up
questions on an answer as one indicator to evaluate the quality of an
answer in the analysis of RQ3.

\begin{figure}[h]
	\centering
	\centering
	\includegraphics[scale=0.20]{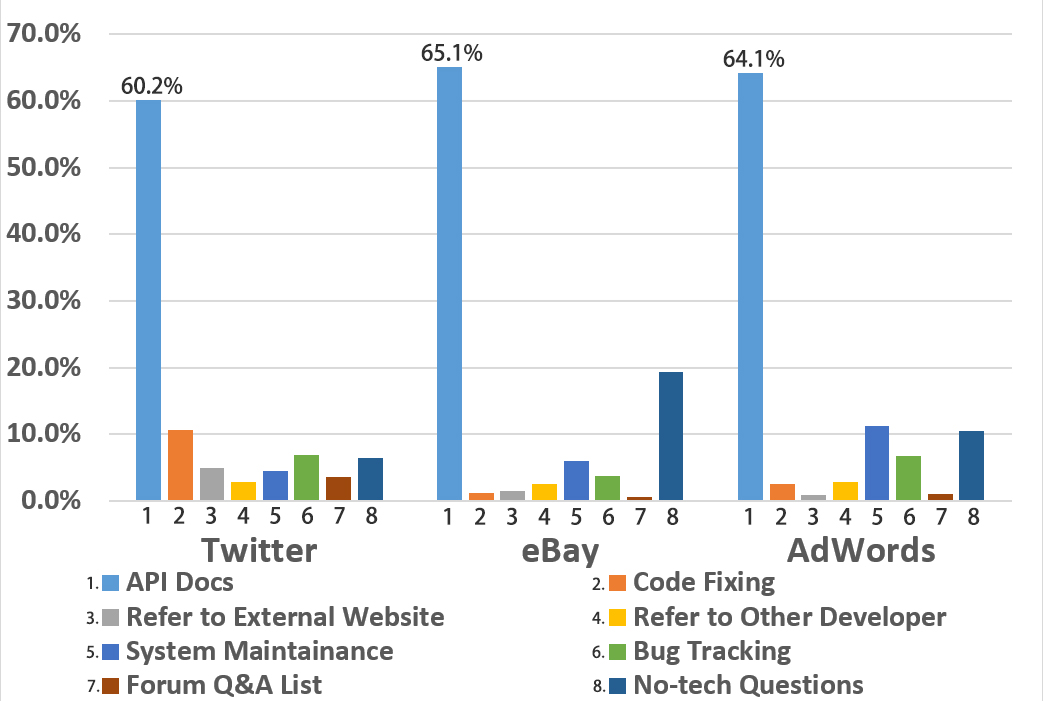}
	%\caption{How questions can be answered}
	\caption{Eight Categories of Questions.}
	\label{questions_classification}
	\vspace{-10pt}
\end{figure}

%\vspace{2pt}
\noindent\textbf{Results of RQs.} Let us present the results of our study.

\textbf{(RQ1.)} The majority of the questions were answered with
providing API method names (or sometimes links to API web
pages). Figure~\ref{questions_classification} shows that we identify 8
categories of questions based on how they were answered. For example,
the \textit{API Docs} indicates the percentage of the studied
questions were answered using an API document, and on average, 63\% of
the questions were answered by using API documentation.  The
\textit{Code Fixing} indicates the percentage of the questions that
were for code errors, and not relevant to the API usages. The
\textit{Refer to External Website} refers to the percentage of the
questions that can not be answered by API documents and need the
information from other external websites.

%and $Refer to Other Developers$ means that the owners of the questions need to contact other developers in the API team or support team.

%8910	5231	14245

%After having these data, we did an initial empirical study by random selecting 368, 358, 374 questions that has been answered in twitter, eBay, and AdWords data set with the confidence level in 0.95 and confidence interval in 0.05. After finish analyzing all these questions, we found many interesting information from them.

\begin{figure}[h]
	\centering
	\begin{subfigure}[b]{0.24\textwidth}
		\centering
		\includegraphics[scale=0.15]{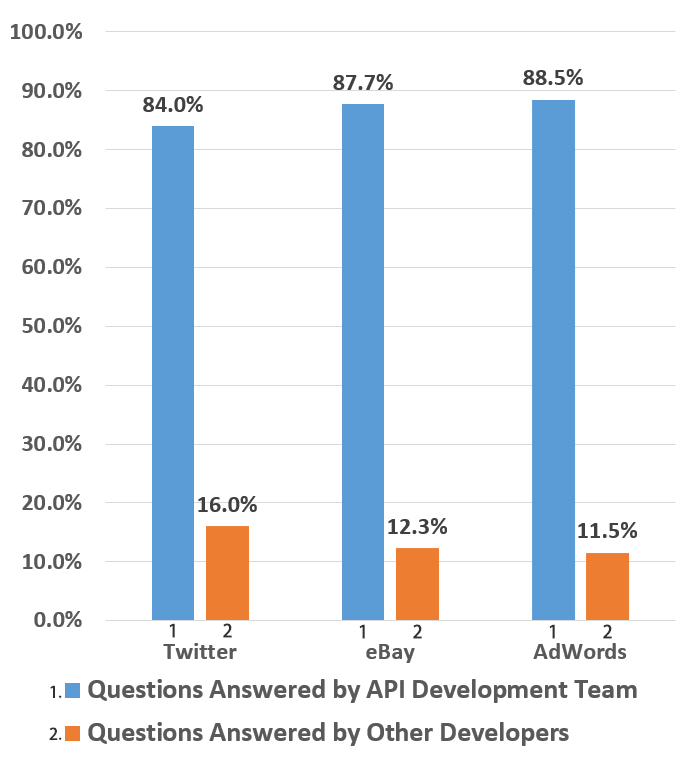}
		\caption{$\;$}
		\label{estudy_results:a}
	\end{subfigure}
	\begin{subfigure}[b]{0.24\textwidth}
		\centering
		\includegraphics[scale=0.15]{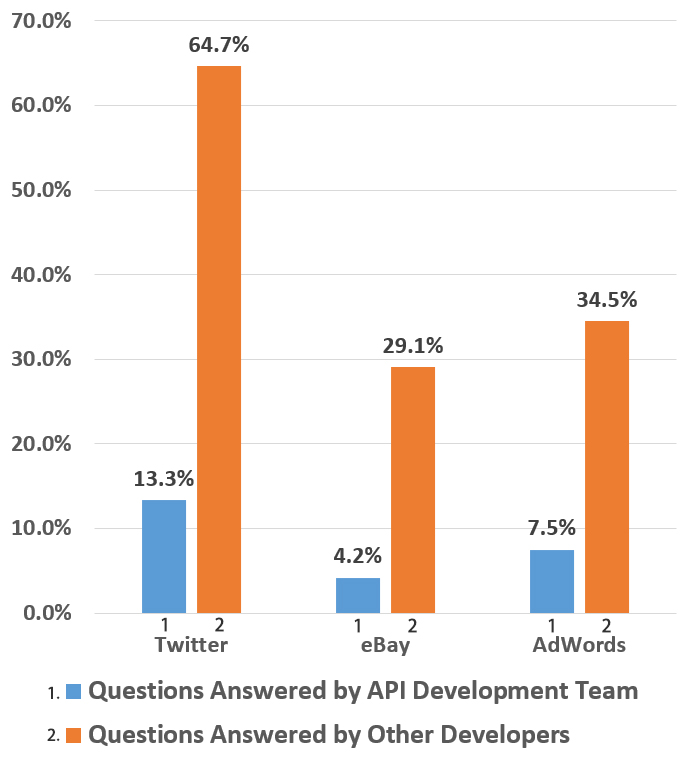}
		\caption{$\;$}
		\label{estudy_results:b}
	\end{subfigure}
	\caption{%Empirical Results of the Analysis on Who Answered the Questions.
		\textbf{(a)} Comparison between the Percentage of Questions Answered by API Development Team and Other Developers. \textbf{(b)} Comparison between the Percentage of Questions Answered by API Development Team and Other Developers, Receiving Follow-up Questions.}
	%There are three different question we analysis in empirical study results: %(a)How questions can be answered (b)Who answered questions (c)How many questions that the developers continue to ask more detailed questions after the question be answered}
	\label{estudy_results}
	\vspace{-10pt}
\end{figure}

\textbf{(RQ2.)} The majority of the studied questions were answered by
API development teams. Figure~\ref{estudy_results:a} shows that
84-87.7\% of the questions were answered by API development teams on
the three forums, while other developers only answered a small
portion of the questions.

\textbf{(RQ3.)} The answers from API development teams have drawn
fewer follow-up questions than the ones answered by other
developers. For example, Figure~\ref{estudy_results:b} shows that only
about 13\% of the questions answered by the Twitter API development
team drew follow-up questions, while about 65\% of the questions
answered by other developers do so. The other API forums share the
same phenomenon.

\section{Discussion}
\vspace{-5pt}
Our three preliminary RQs show that it is important to assist API
development teams to answer developer questions, as on average, +85\%
of the questions were answered by the API development team and fewer
follow-up questions were triggered up after an answer was provided by
an API development team member. Furthermore, on average, +60\% of the
questions were answered using API document links or direct API
method names. Thus, recommending relevant API documents to
answer a question can be very useful.

\section{Threats to Validity}
\vspace{-5pt}
\textbf{Manual Analysis of API forum posts}.  During the labeling
process, most answers can clearly show the relevant APIs. However,
some answers can contain outdated links for API documents, which makes
it very difficult to determine the relevant APIs. We discarded such
answers to try our best to minimize the bias. Our process might bring
bias to our results since we are not familiar with the APIs or from
the development teams.

%Because there is not any existing dataset can show the full meaning and purpose of our approach, we need to build our data sets by ourselves. During the process of doing this, we need to do the labeling manually. There may have different ideas for some labeling between different people, but because of the limitation of time and people amount, we can only do the manual construction of labeled data set by ourselves. Even though most API related questions have clear answers which can help us find the most related API documentations, labeling data by ourselves may still has some interference factors. In order to make our evaluation results become more powerful and convincing, in the future, we would like to invite different people to do the labeling for several times and make sure each labeled data has the best labeling results after discussion with these people.

\textbf{Selection of API forums}.  There are many API forums for
different websites. In our research, we only focused on three very
popular API forums, Twitter, eBay, and AdWords. Thus, we cannot claim
that the result is general for all API forums.

%However, the key drivers of our approach outperforming the baselines are general across forums.

%Applying all these evaluations on different platforms is a time-consuming work, but it is necessary. In the future, we would like to test our approach on more platforms which can test and improve our approach as well.

\section{Related Work}
There has been extensive research devoted to analyzing Stack Overflow, for example, such as analyzing obsolete answers~\cite{zhang2019empirical}, proposing guidelines for writing questions~\cite{calefato2018ask}, discussing best-answer prediction models~\cite{calefato2019empirical}, learning to answer SO questions~\cite{li2018learning}.

\section{Conclusion}
Our empirical results show that it is necessary to build automated
solutions to help the API development teams to answer developers'
questions. We plan to conduct further analysis on API forum posts and
eventually propose solutions to automatically answer developer
questions on API forums.

\section*{Acknowledgments}
This work was supported in part by the US National Science Foundation
(NSF) grants CCF-1723215, CCF-1723432, TWC-1723198, CCF-1518897, and
CNS-1513263.

\bibliographystyle{IEEEtran}
\bibliography{referanceLBR}
\balance
\end{document}